\documentclass[prd,floats,preprint,floatfix,eqsecnum,nofootinbib,12pt]{revtex4}
\pdfoutput=1

\usepackage{amsmath}
\usepackage{graphics, graphicx}
\usepackage{epsfig}
\usepackage{ulem}
\usepackage{color}

\def\1p{{(1p)}}
\def\be{\begin{equation}}
\def\ee{\end{equation}}
\def\beq{\begin{eqnarray}}
\def\eeq{\end{eqnarray}}

\def\p0{\phi_0}
\def\z0{\zeta_0}

\def\Co{C_\ell^{\rm obs}}

\newcommand{\ttle}[1]{}

\begin{document}

\title{Predicting a Prior for Planck}

\author{Thomas Hertog}
\email{thomas.hertog@fys.kuleuven.be}
\affiliation{Institute for Theoretical Physics, KU Leuven, 3001 Leuven, Belgium}

\bibliographystyle{unsrt}

\begin{abstract}

The quantum state of the universe combined with the structure of the landscape potential implies a prior that specifies predictions for observations. We compute the prior for CMB related observables given by the no-boundary wave function (NBWF) in a landscape model that includes a range of inflationary patches representative of relatively simple single-field models. In this landscape the NBWF predicts our classical cosmological background emerges from a region of eternal inflation associated with a plateau-like potential. The spectra of primordial fluctuations on observable scales are characteristic of concave potentials, in excellent agreement with the Planck data. By contrast, alternative theories of initial conditions that strongly favor inflation at high values of the potential are disfavored by observations in this landscape.

\end{abstract}


\maketitle

\section{Introduction}
\label{intro}

The Planck collaboration \cite{Planck13,Planck13b} uses a Bayesian analysis to discriminate between models of inflation on the basis of the Planck satellite data of the temperature variations across the cosmic microwave sky\footnote{See e.g. \cite{Hartle07,Easther12} for other recent applications of Bayesian probability theory to cosmology.}. A Bayesian procedure requires a prior\footnote{Prior probabilities are probabilities that are given independently of any data.}. For their inflationary analysis the Planck collaboration has adopted broad uniform model priors\footnote{Specifically the Planck collaboration considered an ensemble of parameterized models of inflation with a broad range of values of the free parameters in the inflationary potentials (typically varying over several orders of magnitude) and a flat prior on the logarithm of these parameters.}. There is no motivation from theory for this choice of prior. Rather it is designed to reflect ignorance about the underlying theory of cosmology in order to extract information on the physics governing the very early universe implied by data alone. With this prior the Planck collaboration finds strong evidence for an early period of slow-roll inflation driven by a single scalar field. They also find that power law potentials $\lambda \phi^n$ with $n \geq 2$ are disfavored relative to plateau-like potentials \cite{Planck13,Planck13b}.

On the other hand a theory of cosmology that includes both the dynamics and the initial conditions {\it predicts} its own prior. This takes the form of probability distributions for cosmological observables. Observables with distributions that are sharply peaked around specific values are predicted with high accuracy by the theory and can be used to test it against observations. In this paper we mainly focus on the effect of the theory of initial conditions on the prior probabilities. In particular in view of the tight constraints on inflationary theory implied by the Planck observations we aim to distinguish between theories of initial conditions on the basis of their predictions for observables connected to the CMB.

We model the dynamics in terms of a scalar field with a landscape potential that includes a range of inflationary patches that are representative of relatively simple single-field models\footnote{The Planck collaboration treats different models of inflation as different theories. However string theory strongly indicates these should rather be viewed as different patches of a single landscape potential.}. The models differ in their predictions for the scalar spectral tilt $n_s$ and tensor to scalar ratio $r$ of the fluctuations. We assume the landscape itself has no strong statistical bias that (dis)favors a particular inflationary potential within this set of models. This yields an arena where the theory of initial conditions potentially plays an important role in defining the prior.

We model the theory of initial conditions by a semiclassical wave function. A wave function of the universe does not predict a single universe. Rather it populates a given landscape potential with an ensemble of classical universes -- backgrounds {\it and} fluctuations -- and it endows this ensemble with a measure. We adopt the Hartle-Hawking no-boundary wave function (NBWF) \cite{HH83} as a model of the quantum state. At the semiclassical level the NBWF weights universes by $\exp(-I)$ where $I$ is the Euclidean action of a compact regular saddle point that smoothly joins onto the universe when the scale factor is sufficiently large.

It is well-known that the probability distribution over cosmological backgrounds predicted by the NBWF is concentrated around inflationary universes (see e.g. \cite{HHH08,HHH08b}). The NBWF {\it selects} those patches of the landscape where the slow-roll conditions for inflation hold. Hence the landscape essentially reduces to a set of models of inflation in the no-boundary state. The sum of the probabilities of all universes originating in a given inflationary patch of the landscape implies a relative weighting of the `models' of inflation contained in the landscape, which then yields a prior over various observables \cite{HHH10b}.

We are interested in these prior probabilities for local observables connected to the CMB, such as the multipole coefficients $C_l$ characterizing the observed two point correlator. The NBWF predicts the usual probabilities for nearly Gaussian scalar and tensor perturbations around each inflationary background in its ensemble \cite{Halliwell85,HHH10}. Probabilities for observables like the $C_l$'s in a given background follow directly from the Gaussian probabilities for perturbations. The latter depend on the shape of the potential patch probed by the background. This is where the relative weighting of different landscape regions enters: If the NBWF prior is sharply peaked around inflationary backgrounds associated with a particular patch of the landscape then the theory predicts we should observe a CMB spectrum characteristic of the potential in that patch.

It turns out this is the case in our model landscape. We find the NBWF predicts our classical cosmological background emerges from a region of eternal inflation near the maximum of a plateau-like potential in the landscape\footnote{It was conjectured earlier in \cite{HHH08b} that the NBWF favors plateau-like potentials.}. A region of eternal inflation is a very flat patch, associated with scales beyond those directly probed by CMB fluctuations, where $V > \epsilon$, with $\epsilon \equiv V'^2/2V^2$. In a region of eternal inflation the dynamics is governed by the perturbations rather than the classical slow-roll. In plateau-like potentials the eternal inflation condition $V > \epsilon$ generally holds near the maximum.

The spectral features of the primordial fluctuations on observable scales are predicted to have the usual characteristics of inflationary fluctuations in plateau-like potentials. In particular the theory predicts the combination of $n_s$ and $r$ to be in the region of the $(n_s,r)$-plane corresponding to concave potentials (cf \cite{Planck13}), in excellent agreement with the Planck data. Finally we find evidence that among the plateau potentials those with a lower maximum are more strongly favored. Within the context of our model landscape this leads to the further prediction of a low tensor to scalar ratio $r$ in the CMB fluctuations. 

These predictions rely not only on the structure of the landscape but also on the no-boundary theory of initial conditions. In particular, had we done our analysis in the same landscape model but with a wave function that strongly favors inflation at high values of the potential we would have found a rather different result. We return to this point in the conclusion.

\section{A Landscape Model}
\label{landscape}

String theory appears to give rise to a landscape of possible worlds. The string landscape is thought to contain a vast number of vacua, including some that have four large dimensions, a small positive cosmological constant $\Lambda$, and the Standard Model. In this paper we condition on these features and focus on the shape of the scalar potential in the neighborhood of such vacua. To this end we consider a toy model version of the landscape given by a scalar $\vec \phi$ with a multidimensional potential $V(\vec \phi)$ with a single minimum where $\Lambda$ takes the observed value, and with a range of different potential behaviors in different directions in field space. 

There is increasing evidence that the landscape contains a wide variety of potential patches where the slow-roll conditions for inflation hold\footnote{For a pedagogical review of inflation in string theory see \cite{Baumann09}.} including regions that admit large-field inflation \cite{Silverstein08,McAllister10}. On the other hand there is at present no indication that the landscape itself exhibits an exponentially strong statistical bias (dis)favoring a particular inflationary potential\footnote{It would be interesting to generalize our analysis to a sector of the string landscape where the relative frequency of different inflationary potentials can be reliably computed (see e.g. \cite{Argawal11,Westphal13} for recent work on this). The predictions for observations will then be governed by a combination of the no-boundary weighting and the specific structure of the landscape potential.}. 

We incorporate these features of the string landscape in our toy model by including directions in field space in which the conditions for inflation are satisfied. We assume different inflationary directions of $V$ are separated by steep barriers and that each direction can be described in terms of an effective single-field model. We consider a number of families of models that are representative of single-field potentials with one or two free parameters\footnote{The restriction to a small number of free parameters excludes landscape directions where the potential is of the false vacuum form \cite{Freivogel06}. Such directions give rise to a contribution from open inflationary universes to the NBWF \cite{Gratton00}. However false vacua at high potential values turn out to be suppressed by the NBWF prior. False vacua at low values require more parameters compared to the potentials we consider, and therefore plausibly occur much less frequently in the landscape. Hence we do not expect that including false vacuum directions will significantly change our results.}. Specifically our analysis covers exponential and power law potentials \cite{Linde83,Silverstein08,McAllister10} as well as plateau-like potentials such as Starobinsky's original model of $R^2$ inflation \cite{Starobinsky80} in the Einstein frame and the symmetry-breaking potential \cite{Olive90} leading to new inflation \cite{Albrecht82,Linde82}. The models we consider differ in their predictions for the scalar spectral tilt $n_s$ and tensor to scalar ratio $r$ of the fluctuations. On the other hand we restrict attention to models where $\Lambda$ and the overall amplitude of the primordial fluctuations take their observed values\footnote{The generalization of our analysis to landscape models in which also $\Lambda$ and the overall amplitude of the primordial fluctuations varies is discussed elsewhere \cite{HH13}.}. 

Most of these models admit a regime of eternal inflation associated with scales beyond those directly probed by CMB fluctuations. We assume the landscape itself has no strong statistical bias that (dis)favors a particular inflationary potential within this set of models. Hence we include no additional weighting factor to account for the relative frequency with which different inflationary potentials occur in the landscape. This will turn out to be a good approximation provided the statistical bias is not exponentially strong.

\begin{figure}[t]
\includegraphics[width=3in]{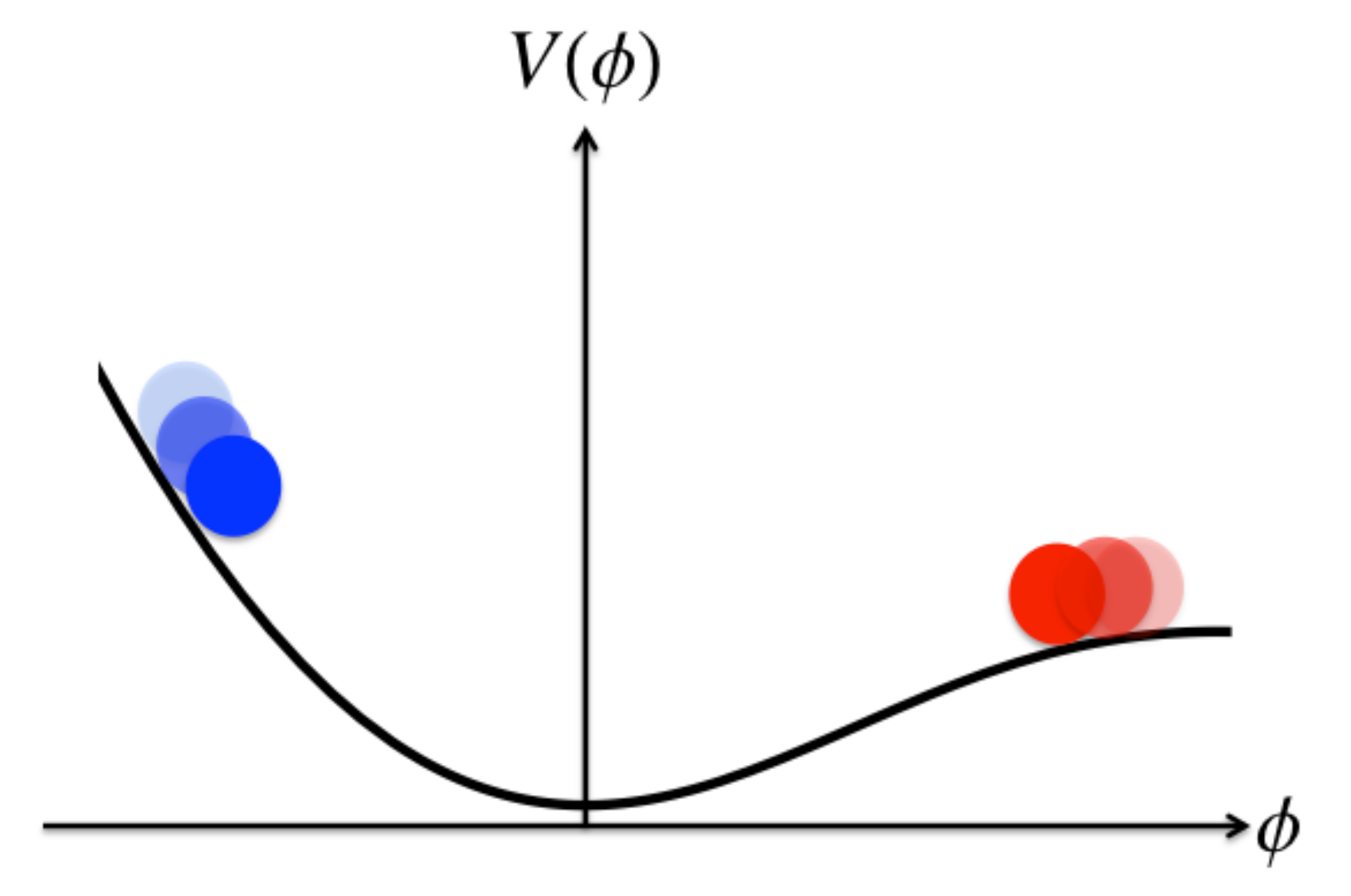}
\caption{A slice through the toy model landscape. We are at the bottom of $V$ at the present time. The pattern of fluctuations in the microwave sky that we observe today depends on the direction in field space the early inflationary universe rolled down from. A theory of initial conditions in the form of a wave function of the universe yields a prior that predicts which spectrum we should see. }
\label{slice}
\end{figure}

In this toy landscape we find ourselves near the minimum of the potential at the present time. If the field was displaced from the minimum in the past then our universe may have undergone a period of scalar field driven inflation. The pattern of fluctuations in the microwave sky that we observe today will then depend on which direction in field space the early inflationary universe rolled down from (cf. Fig \ref{slice}). It is a central goal of theoretical cosmology to predict the pattern we should see. The structure of the landscape is not sufficient to determine this. Indeed the landscape alone does not specify whether inflation occurred at all. To proceed we must first complete the theory with a model of the quantum state of the universe.

\section{No-Boundary Measure}
\label{measure}

At the semiclassical level a quantum state is defined in terms of a wave function $\Psi[h_{ij},\phi]$ on the superspace of real geometries $h_{ij}(x)$ and scalar field configurations $\phi(x)$ on a spacelike three-surface $\Sigma$ that we take to be closed. We adopt the semiclassical no-boundary wave function (NBWF) \cite{HH83} as a model of this state. The NBWF is a natural choice of state, analogous e.g. to the ground state in quantum mechanics, with a strong motivation in fundamental theory\footnote{At the semiclassical level the NBWF is specified by a regularity condition that selects the saddle points contributing to the wave function. A similar boundary condition of regularity is universally implemented in the context of the Euclidean version of AdS/CFT, which can be thought of as calculating the wave function of a Euclidean universe in the large volume limit \cite{Horowitz04}. The saddle points of the NBWF associated with Lorentzian universes can be viewed as complex versions of such Euclidean AdS domain walls \cite{HH11}.}. The semiclassical NBWF weights different configurations $(h_{ij},\phi)$ by $\exp(-I)$ where $I$ is the Euclidean action\footnote{We use Planck units where $\hbar=c=8\pi G=1$.}  of the dominant compact, regular saddle point solution that matches $(h_{ij},\phi)$ on its only boundary $\Sigma$. When the surfaces $\Sigma$ are three spheres of sufficiently large radius $a$ the dominant saddle points are complex. Then one gets \cite{HHH08}
\begin{equation}
\Psi[h_{ij},\phi] \approx  \exp(-I[h_{ij},\phi]) = \exp(-I_R[h_{ij},\phi] +i S[h_{ij},\phi])
\label{semiclass}
\end{equation}
where $I_R[h_{ij},\phi]$ and $-S[h_{ij},\phi]$ are the real and imaginary parts of the Euclidean action $I$ evaluated at the extremum. If $S$ varies rapidly compared to $I_R$  the wave function takes a WKB form and predicts the configuration evolves as a real classical universe, according to the Lorentzian Einstein equations. Vice versa, a given classical universe is predicted by the NBWF only if its geometry and field history can be associated with a regular saddle point for which \eqref{semiclass} has a rapidly varying phase. 

In many dynamical models the NBWF predicts a multiverse consisting of a family of classical, closed Lorentzian universes. The individual universes in this multiverse are the integral curves of $S$ and have tree-level probabilities proportional to  $\exp[-2 I_R(h_{ij},\phi)]$, which is constant along the integral curve \cite{HHH08}. Given that all our observations are conditioned on the existence of classical space-time we focus on the classical predictions of the NBWF from here onwards.

In our toy model landscape these include an ensemble of Lorentzian Friedman-Lema\^itre-Robertson-Walker (FLRW) backgrounds with Gaussian perturbations \cite{HHH08,HHH10,Hertog13}. The NBWF has the striking property that it predicts only classical FLRW backgrounds which undergo some amount of matter driven slow-roll inflation. Intuitively this is because only universes with initially sufficiently small gradients can be smoothly glued onto regular saddle points. Hence the NBWF populates the landscape in a very specific manner: it provides a prior that {\it selects} landscape patches where the conditions for inflation hold. 

The model landscape we consider has only single-field inflationary directions. In each of these the NBWF selects a one-parameter set of inflationary backgrounds. We label a background in direction $J$ by the absolute value $\phi_0^J$ of the scalar field  at the `South Pole'  (SP) of its corresponding saddle point, where $a \rightarrow 0$. The value $\phi_0^J$ is approximately equal to the largest value of the scalar field in the corresponding classical background. Hence the number of efolds $N(\phi_0^J) \approx \int_{\phi_e^J}^{\phi_0^J} (V/V')$ where $\phi_e^J$ is the value at which $\epsilon^J=1$ and hence inflation ends. The range of $\phi_0^J$ also has a lower bound at a critical value $\phi_{0c}^J$, with $N(\phi_{0c}^J) \sim {\cal O}(1)$, below which the NBWF predicts no classical cosmological backgrounds \cite{HHH08}.

The probability distribution $P(\phi_0^J)$ over backgrounds is proportional to \cite{HHH08}
\be
\exp[-2 I_R] \approx \exp [3\pi/V(\phi_0^J)] 
\label{bu}
\ee
The probabilities for scalar curvature perturbations $\zeta^J$ and tensor perturbations $t_{ij}^J$ can be obtained from the semiclassical wave function of perturbations around homogeneous saddle points. They are given by the usual product\footnote{The NBWF is the cosmological analog of the ground state; the regularity condition on the saddle points implies that perturbations start out approximately in the Bunch-Davies ground state \cite{Halliwell85}.} of Gaussian probabilities $P(\zeta_k^J|\phi_0^J)$ and $P(t_{ij(k)}^J|\phi_0^J)$ for modes $\zeta_k^J,t_{ij(k)}^J$ on $S^3$ \cite{Halliwell85,HHH10}. In particular the NBWF for $\zeta^J$ implies the usual scalar power spectrum $\Delta^2_{\zeta^J} \equiv (k^3/2\pi^2) {\cal P}^J_{\zeta}(k) = A^J_s (k/k_{*})^{n^J_s-1}$ where $A^J_s \sim (V/\epsilon)^J$ and $n^J_s-1 = 2 \eta^J - 6\epsilon^J$, with $\eta^J \equiv (V''/V)^J$, are evaluated at the value $\phi_{*}$ where the reference scale $k_{*}$ crosses the horizon during inflation, and similarly one gets the usual spectrum of tensor perturbations. Observables associated with the CMB can be expressed in terms of $\zeta^J$ and $t_{ij}^J$, and their probability distributions follow from $P(\zeta_k^J|\phi_0^J)$ and $P(t_{ij(k)}^J|\phi_0^J)$. 

For perturbations on currently observable scales it is well-known that $\Delta_{\zeta}^2 \sim {\cal O}(10^{-9})$. 
Saddle points with $V \ll \epsilon$ at the SP have $V \ll \epsilon$ everywhere and predict an ensemble of nearly homogeneous inflationary universes with small fluctuations. By contrast, saddle points starting in regions of the landscape where $V > \epsilon$ predict high amplitudes for significantly inhomogeneous universes that have large (very) long-wavelength perturbations on scales associated with this region. This is because a landscape region where $V > \epsilon$ corresponds to a regime of eternal inflation, where the typical size of perturbations on the horizon scale is comparable or larger than the classical field motion in a Hubble time. As a consequence the perturbations have a large effect on those scales. We denote the threshold field value for which $V =\epsilon$ at the SP by $\phi_{EI}^J$. In our model landscape, the NBWF ensemble of universes associated with a given direction generally contains saddle points corresponding to nearly homogeneous universes as well as eternally inflating histories that are inhomogeneous on the largest scales.

\section{Probabilities for Observations}
\label{pred}

Predictions for observations follow from conditional probability distributions $P({\cal O}|D)$ for different values of a cosmological observable ${\cal O}$ given the existence of (at least one instance of) our observational situation $D$ \cite{HHH08b,HH09}. The condition on the observational situation is the more general analog in cosmology of the experimental setup in the lab. The distributions $P({\cal O}|D)$ are obtained from the probability distributions \eqref{bu} over the classical histories of the universe by conditioning on $D$, taking in account its possible locations in each history, and summing over features that are unobserved \cite{HHH10b}. In our model landscape the conditional distributions take the form
\be
\label{Pobs}
P({\cal O}|D) \propto \sum_J \int P({\cal O}|\phi_0^J,\zeta_{k}^J) P(D|{\cal O},\phi_0^J, \zeta_{k}^J) P(\phi_0^J, \zeta_{k}^J)
\ee
where the integral is over $\phi_0^J$ and $\zeta_{k}^J$ labeling the ensemble of classical histories (backgrounds and perturbations) in the different potential direction $J$. The last factor is the NBWF prior given by \eqref{bu} together with the probabilities $P(\zeta_k^J|\phi_0^J)$ for scalar perturbations\footnote{For simplicity we coarse grain over the tensor perturbations in the remainder of this paper.}.

The `top-down' (TD) requirement\footnote{See \cite{HH06} for the origin of this terminology. We refer to \cite{HH09} for a more detailed discussion of the top-down factor $P(D|{\cal O},\phi_0^J, \zeta_{k}^J)$ in \eqref{Pobs}.} that our observational situation $D$ exists somewhere is of course a trivial condition in those histories in the ensemble that are sufficiently large. By contrast, in members $(\phi_0^J, \zeta_{k}^J)$ of the classical ensemble that are small one expects the TD factor $P(D|{\cal O},\phi_0^J, \zeta_{k}^J)$ in \eqref{Pobs} will be exceedingly small for realistic $D$. This is because as observers we are physical systems within the universe, described by local data $D$, which occur only with a small probability $p_E(D)$ in any Hubble volume on a surface $\Sigma$ of approximately constant density. Specifically in classical histories with $N_h \ll 1/p_E(D)$, where $N_h(\p0^J,\zeta^J)$ is the total number of Hubble volumes in $\Sigma$, the TD factor $P(D|{\cal O},\phi_0^J, \zeta_{k}^J) \approx p_E N_h \ll 1$ \cite{HH09}. Hence the TD factor leads to a weighting of the `bottom-up' (BU) probabilities $P(\phi_0^J, \zeta_{k}^J)$ by the volume of $\Sigma$ in this regime of the distribution\cite{Page97,HHH08b,HH09}\footnote{This should not be confused with the volume weighting that features in certain classical studies of the measure problem in eternal inflation. There volume weighting is a defining property of certain measures. By contrast in quantum cosmology the measure is supplied by the wave function of the universe, and volume weighting emerges only in a regime of the ensemble of histories {\it without} eternal inflation. It should be views as a property of a particular class of conditional probabilities calculated in the quantum mechanical ensemble of histories predicted by the universe's quantum state.}. We have argued \cite{HH09} this is the case in saddle point histories that start below the threshold $\phi_{EI}^J$ of eternal inflation, which predict high amplitudes for nearly homogeneous universes only, with $Vol(\Sigma) \propto e^{3N}$. 
A more accurate description of the observational situation employing a larger set of our available data $D$ decreases $p_E$ and thus more strongly suppresses this class of histories in the ensemble.

By contrast, saddle points corresponding to histories with a regime of eternal inflation predict high amplitudes for configurations that have large perturbations $\zeta_{k,lw}^J$ on scales associated with eternal inflation \cite{HHH10}. As a consequence the volume of $\Sigma$ of such histories is generally exceedingly large or even infinite so that $P(D|{\cal O},\phi_0^J, \zeta_{k}^J) \approx 1$ when $\phi_0^J \geq \phi_{EI}^J$. In potential directions with a regime of eternal inflation the TD factor therefore takes the form of a step function\footnote{We are assuming $p_E(D) >0$. This is the case in our model landscape, in which both $\Lambda$ and the overall amplitude of the primordial fluctuations are kept fixed at their observed values. For a more general discussion see \cite{HH13}.}, rising rapidly from $p_E N_h \ll 1$ to nearly one. Hence in the neighborhood of the threshold of eternal inflation and for $\phi_0^J > \phi_{EI}^J$ we have\footnote{We have included a factor $3/2$ in the argument of the $\tanh$ so that the expression reproduces the volume weighting $\sim e^{3N}$ for $\phi_0^J$ well below $\phi_{EI}^J$, albeit for the particular case $p_E = \exp(-3N(\phi_{EI}^J))$.},
\be
P(D| {\cal O},\phi_0^J, \zeta_{k}^J) \approx  \frac{1}{2}\left[1+\tanh [ \frac{3}{2} (N(\phi_0^J) - N(\phi_{EI}^J)]\right] 
\label{td}
\ee

Finally there are the probabilities $P({\cal O}|\phi_0^J,\zeta_{k}^J)$ for the local observable ${\cal O}$ of interest. Here we concentrate on observables associated with the CMB. These are given by a function $F_{\cal O}(\zeta^J,\phi_0^J)$ that involves only small, short wavelength fluctuations $\zeta_{k,sw}^J$ that can be observed in one Hubble volume. In particular they are independent of the detailed structure of $\Sigma$ on scales far beyond our horizon \cite{Giddings11}. This means the long-wavelength fluctuations $\zeta_{k,lw}^J$ can be coarse grained over in the calculation of $P({\cal O}|\phi_0^J,\zeta_{k}^J)$. To leading order in $\hbar$, if one coarse grains the probabilities over all possible values of the long-wavelength fluctuations this sum-over-histories yields one \cite{HHH10b}. Integrating over $\zeta_{k,lw}^J$, taking in account their effect on the TD factor, and substituting \eqref{bu} in \eqref{Pobs} yields
\be
\label{Pobs2}
P({\cal O}|D) \propto \sum_J \int P({\cal O}|\phi_0^J,\zeta_{k,sw}^J)P(\zeta_{k,sw}^J|\phi_0^J) P_{TD}(\phi_0^J,\phi^J_{EI},p_E)
\exp [3\pi/V(\phi_0^J)] 
\ee
where the integral is now over $\phi_0^J$ and $\zeta_{k,sw}^J$. As discussed above the TD factor $P_{TD}$ is approximately given by $p_E N_h \ll 1$ for $\phi_0^J < \phi_{EI}^J$ and by a step function of the form \eqref{td} for $\phi_0^J \sim \phi_{EI}^J$. 

Hence the probabilities specifying the predictions for our observations can be estimated using saddle points that are nearly homogeneous everywhere and retain only the small, short-wavelength, observable, fluctuations\footnote{The distribution over nearly homogeneous saddle points in \eqref{Pobs2} arises as the result of a coarse-graining over long-wavelength perturbations. Therefore the contribution of each homogeneous saddle point in this distribution should be interpreted as an estimate of the sum of the probabilities of an ensemble of perturbed, classical histories with widely different structures on scales associated with eternal inflation. In particular the approximate homogeneity of the saddle points constitutes neither a prediction nor an assumption of homogeneity on those scales.}. The probability distribution over coarse grained backgrounds in \eqref{Pobs2} yields a relative weighting of landscape directions and of inflationary backgrounds within a given direction. This is the central contribution of a theory of the quantum state to inflationary cosmology.

The TD weighting in \eqref{Pobs2} has a significant effect on the prior. The BU probabilities \eqref{bu} favor backgrounds starting at a low value of the potential followed by only a few efolds of slow-roll inflation. However, the TD weighting strongly suppresses the contribution from universes starting below the threshold of eternal inflation. Instead it favors saddle points in regions of eternal inflation, simply because there are more possible locations of our past light cone on the resulting surfaces $\Sigma_s$. In the next section we show that in directions of the landscape in which the potential has a regime of eternal inflation the low BU probability of histories starting above the threshold for eternal inflation is more than compensated for by the TD weighting.

\section{NBWF Predictions in Landscape Models}
\label{pred}

We now look in more detail at the contributions to \eqref{Pobs2} from different landscape directions. We first consider directions in which the potential takes a power law form $V(\phi) = \lambda \phi^n$. In this class of potentials inflation can occur for $\phi \gg 1$ and ends when the slow-roll conditions break down at $\phi_e \approx n/\sqrt{2}$. In each power law direction the NBWF predicts a one-parameter family of inflationary backgrounds which can be labeled by $\phi_0 \geq \phi_{0c}$ where $\phi_0$ is the absolute value of $\phi$ at the SP of the corresponding saddle point and the lower bound $\phi_{0c} \sim {\cal O}(n)$. The number of efolds is given by $N \approx \phi_0^2/2n$. The threshold $V \sim \epsilon$ of eternal inflation is given by $\phi_{EI} \approx (n^2/\lambda)^{1/n+2}$. The amplitude $A_s^{*} = (V/\epsilon)^{*}$ of the scalar curvature power spectrum $\Delta_{\zeta}^2$ on COBE scales $k_{*}$ determines $\lambda =  n^2 A_s^{*}/2(2nN_{*})^{n+2/2}$ where $N_{*} \sim {\cal O}(50)$ is the number of efolds before the end of inflation at the value $\phi_{*}$ where the COBE scale $k_{*}$ leaves the horizon. For $n=2$ the COBE normalization $A_s \sim 10^{-9}$ implies $\lambda \sim 10^{-12}$.

\begin{figure}[t]
\includegraphics[width=2.7in]{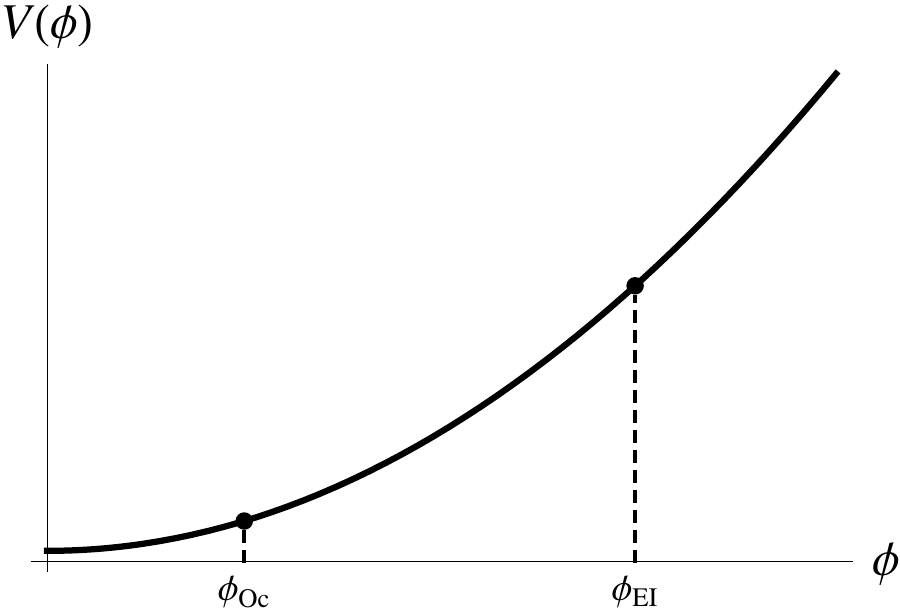}\hfill
\includegraphics[width=3in]{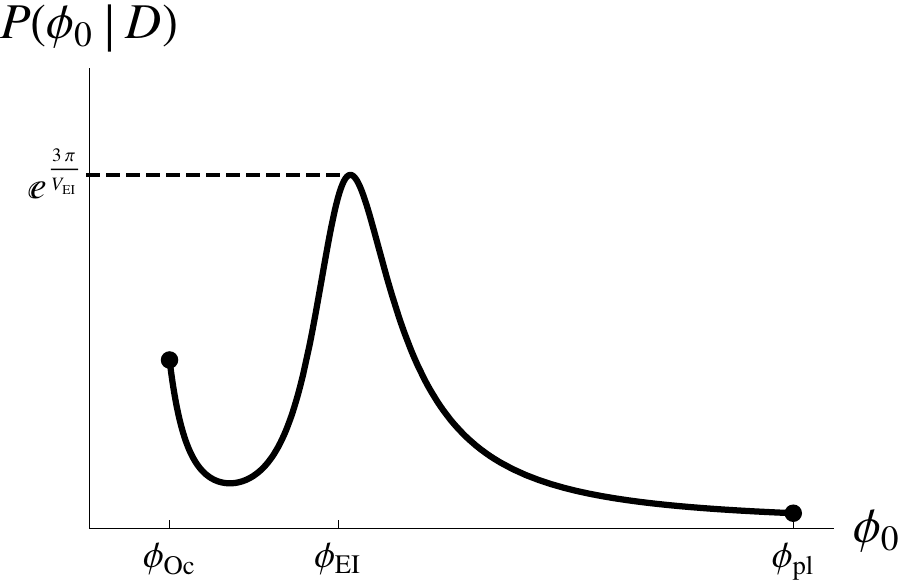}
\caption{The NBWF probability distribution over backgrounds \eqref{plot} conditioned on the existence of our observational situation $D$, with $p_E$ given in the text, in a quadratic potential with $m^2 \sim {\cal O}(10^{-2})$. The eternal inflation threshold $\phi_{EI} \sim 1/\sqrt{m}$, well below the Planck density at $\phi_{pl} \sim 1/m$, and the lower bound from classicality $\phi_{0c} =1.4$. All backgrounds have some amount of inflation with $N \sim \phi_0^2$. The peak of the distribution around $\phi_{EI}$ is a universal feature of the top-down probabilities over backgrounds in power law potentials.}
\label{mono}
\end{figure}

For power law directions the dominant contribution to the integral over $\phi_0^J$ in \eqref{Pobs2} comes, for sufficiently small $p_E$, from saddle points with a regime of eternal inflation. To illustrate this we plot in Fig \ref{mono} the distribution over backgrounds 
\be
P_{bkgd} \equiv P_{TD}(\phi_0^J,\phi^J_{EI},p_E) \exp [3\pi/V(\phi_0^J)] 
\label{plot}
\ee
that enters in \eqref{Pobs2}, for the case $p_E = \exp(-3N(\phi_{EI}^J))$ (for which the TD factor equals \eqref{td} for all $\phi_0$.). Fig \ref{mono} shows this for a quadratic potential with $\lambda \sim {\cal O}(10^{-2})$ but the qualitative behavior of the distribution shown in Fig \ref{mono} is representative for all power law potentials with a regime of eternal inflation. 

The origin of both peaks in this distribution can easily be understood. For $\phi_0 \ll \phi_{EI}$ the histories are nearly homogeneous, so the TD factor in \eqref{Pobs2} is $p_E N_h = p_E e^{3\phi_0^2/2n}$ and does not compete with the bottom-up NBWF weighting on histories. The distribution is therefore tilted towards $\phi_{0c}$ in this regime, with $P_{bkgd}(\phi^J_{0c}) \approx p_E \exp [3\pi/V(\phi_{0c}^J)]$. By contrast for $\phi_0 \geq \phi_{EI}$ even the `homogeneous' TD factor $\propto  e^{3\phi_0^2/2n}$ more than compensates the NBWF weighting leading to a tilt of the distribution towards large $\phi_0$ \cite{HH09}. The step function form \eqref{td} of the TD factor, which takes in account the back reaction of large-scale inhomogeneities on the volume of constant density surfaces, further enhances this effect. Beyond the eternal inflation threshold the TD factor \eqref{td} equals one, so the distribution in this regime is again governed by the NBWF weighing on histories and therefore tilted towards $\phi_0 \sim \phi_{EI}$. Together this leads to a second peak around $\phi_0 \sim \phi_{EI}$, with $P_{bkgd}(\phi^J_{EI}) \approx \exp [3\pi/V(\phi_{EI}^J)]$. The relative height of both peaks is therefore determined by the shape of $V$ and by the value of $p_E$. Specifically a more accurate characterization of our observational situation employing a larger set of our available data $D$ decreases $p_E(D)$ and suppresses the first peak if $p_E < \exp[3\pi/V(\phi_{EI}^J) -3\pi/V(\phi_{0c}^J)]$. It was shown in \cite{HH09} that realistic values of $p_E$ based on only a very small part of our available data obey this condition and therefore suffice to ensure the second peak at $\phi_0 \sim \phi_{EI}$ dominates the distribution. As mentioned above for the figures we have used $p_E = \exp(-3N(\phi_{EI}^J))$ for which the form of the TD factor simplifies.

For realistic values of $\lambda \ll 1$ -- which we assume in this paper -- and for sufficiently small $p_E$ the distribution is sharply peaked around $\phi_{EI}$. In what follows we therefore approximate the integral over $\phi_0^J$ in power law directions by selecting the background with $\phi_0^J = \phi_{EI}^J$. This also implies that power law directions without a regime of eternal inflation\footnote{This is the case e.g. when corrections steepen the potential at large $\phi$.} are strongly suppressed relative to directions with eternal inflation. For the same reason other inflationary landscape directions without eternal inflation, such as exponential potentials, provide a negligible contribution to the probabilities \eqref{Pobs2}.

Next we consider plateau-like potentials of the form $V(\phi) = V_0(1 - \phi^n/\mu + \cdots)$ during inflation. For $n=2$ slow-roll inflation requires $\mu \gg 1$ and necessarily involves a large field range. For $n \geq 3$ these include both large-field and small-field models. The NBWF predicts a one-parameter family of backgrounds in plateau-like direction with a region where the slow-roll conditions hold. Inflation ends when $\phi = \phi_e \approx (2\mu/n)^{1/n-1}$. The number of efolds is $N \approx \mu \phi_0^{2-n}/n(n-2)$ for $n \geq 3$ and $N \approx (\mu/2) \log(\phi_e/\phi_0)$ for $n=2$.
The condition $V > \epsilon$ for eternal inflation holds for $\phi < \phi_{EI}= (\mu^2 V_0/n^2)^{1/2n-2}$. 
Hence saddle point histories with $\phi_0 >\phi_{EI}$ are nearly homogeneous whereas histories with $\phi_0 <\phi_{EI}$ are inhomogeneous on the largest scales. In terms of the potential parameters the scalar amplitude on COBE scales $A_s^{*} = 2(n-2)N_{*}V_0/n(\phi_{*}^n/\mu)$ for $n \geq 3$ and $A_s^{*} \approx V_0 e^{2N_{*}/\mu}$ for $n=2$.

\begin{figure}[t]
\includegraphics[width=2.7in]{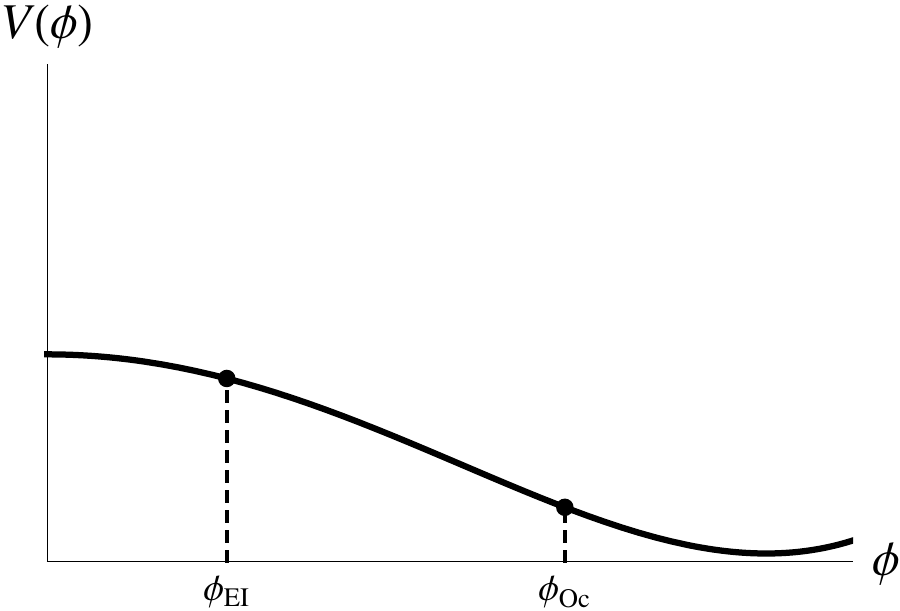}\hfill
\includegraphics[width=3in]{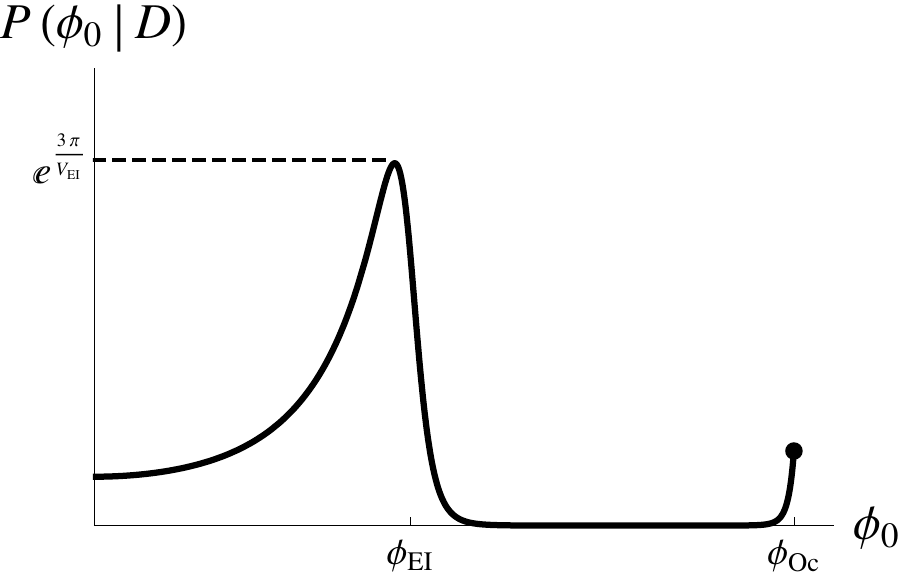}
\caption{The NBWF probability distribution over backgrounds \eqref{plot} conditioned on the existence of our observational situation $D$ in a hill-top potential of the form $V(\phi) = V_0(1 - \phi^2/\mu + \cdots)$ during inflation, with $\mu \sim {\cal O}(10^{-2})$. The shape of the distribution with its peak around $\phi_0 \sim \phi_{EI}$ is characteristic of a large class of plateau-like potentials.}
\label{plat}
\end{figure}

In Figure \ref{plat} we plot the distribution $P(\phi_0|D)$ over backgrounds entering in \eqref{Pobs2} for a hill-top potential with $n=2$.
The dominant contribution comes again from saddle points with a regime of eternal inflation, for the same reasons as discussed above. As before the case shown is representative for a broad class of plateau-like potentials including the $n \geq 3$ hill-top models and $R^2$ inflation \cite{Starobinsky80} which in the Einstein frame has a potential\footnote{$R^2$ inflation can be viewed as the large $n$ limit of hill-top models.}
 $V(\phi) = V_0(1-e^{-\sqrt{2/3}\phi})$. For realistic values of the parameters we find the probabilities are very sharply peaked around $\phi_{EI}$. Hence we approximate the integral over $\phi_0^J$ in plateau-like landscape directions by selecting the background with $\phi_0^J = \phi_{EI}^J$.  
 
Taken together this means the NBWF {\it selects} the landscape directions exhibiting a regime of eternal inflation. Thus eq. \eqref{Pobs2} reduces to
\be
P({\cal O}|D) \approx \sum_{J_{EI}^{\ }} \exp [3\pi/V(\phi_{EI}^J)] \int_{\zeta_{k,sw}^J} P({\cal O}|\phi_{EI}^J,\zeta_{k,sw}^J) P(\zeta_{k,sw}^J|\phi_{EI}^J)
\label{res}
\ee
where $J_{EI}^{\ }$ labels the directions with eternal inflation. We derived this result in the context of a specific model landscape comprising single-field power law and plateau-like potentials. However the TD weighting \eqref{td} is universal. It is plausible therefore that regions of eternal inflation dominate the probabilities in a much broader class of landscape models, including landscapes with false vacua potentials and multi field patches, and for a wide range of quantum states.

The factor $P({\cal O}|\phi_{EI}^J,\zeta_{k,sw}^J)$ in \eqref{res} that gives the probabilities for local observables ${\cal O}$ associated with perturbations $\zeta_{k,sw}^J$ in a given background, labeled by $\phi_{EI}^J$, is given by
\be
\label{fluct}
\int_{\zeta_{k,sw}}  \delta({\cal O} - F_{{\cal O}}(\phi^J_{EI},\zeta^J_{k,sw})) P(\zeta^J_{k,sw}|\phi^J_{EI}) \equiv P({\cal O}|\phi_{EI}^J)
\ee
where $F_{\cal O}$ is a function that expresses ${\cal O}$ in terms of $\zeta_{k,sw}$. As an example consider the standard multipole coefficients $\Co$ of the observed CMB two point correlator. The NBWF probabilities for fluctuations around a given background are Gaussian to lowest order in their amplitude. The probabilities $P^\ell(\Co|D)$ for a given $\ell$ are therefore essentially a $\chi^2$-distribution specified by a mean $\langle \Co \rangle = C_l^J $ and (cosmic) variance  $\sigma^J_\ell \equiv 2(C_l^J)^2/(2\ell+1)$ where the $C_l^J$ are the theoretical multipole coefficients that completely characterize the Gaussian fluctuations in direction $J$.  

The distribution \eqref{res} selects the direction(s) where the threshold for eternal inflation lies at the lowest value of the potential, independently of its shape above this value. The dominant effect on the distribution \eqref{res} in our model landscape comes from the smallness of the COBE amplitude $A_s^{*}$ of the scalar curvature spectrum. In terms of $A^{*}_s$ the background factor in \eqref{res} for power law potentials reads 
\be
\exp [3\pi/V(\phi_{EI})]  = \exp\left[\frac{12\pi N_{*}}{n(A_s^{*})^{2/2+n}}\right]
\label{pw}
\ee
whereas for the hill-top potentials we have
\be
\exp [3\pi/V(\phi_{EI})]  = \exp\left[\frac{6\pi(n-2) N_{*}}{nA_s^{*}(\phi_{*}^n/\mu)}\right]
\label{hilltop}
\ee
if $n \geq 3$ and $\exp\left[3\pi e^{2N_{*}/\mu}/A_s^{*}\right]$ for $n=2$. For $R^2$ inflation we get  $\exp\left[4\pi N_{*}^2/A_s^{*}\right]$. 

A comparison of \eqref{pw} and \eqref{hilltop} shows $A_s^{*}$ enters with a different power, strongly favoring plateau-like potentials over power law potentials. {\it The NBWF prior thus predicts that our universe emerged from a region of eternal inflation associated with a plateau-like potential in the landscape.}

Eqs \eqref{res} and \eqref{fluct} imply that the dominant set of backgrounds determine the predictions for the spectral features of the primordial fluctuations we should expect to observe. In our model landscape the leading order fluctuation spectra associated with different directions differ in their predictions for the scalar tilt $n_s$ and the tensor to scalar ratio $r$ on observable scales. Since the dominant plateau-like potentials are everywhere concave we predict the combination of $n_s$ and $r$ to be in the region of the $(n_s,r)$-plane corresponding to concave potentials (cf Fig 1 in \cite{Planck13}).

Among the different plateau potentials the NBWF favors $R^2$ inflation and those models where the ratio $\phi_{*}^n/\mu$ takes a low value. Since $\epsilon \sim \phi_{*}^n/(N_{*}\mu)$ this leads to the further prediction of a low tensor to scalar ratio in the fluctuations. The NBWF prior therefore most strongly favors those plateau models that lie near the horizontal axis in the celebrated $r$ versus $n_s$ diagram (cf Figure 1 in \cite{Planck13}). 

It is instructive to compare our results with those of the Planck collaboration, which also found that plateau-like potentials are favored relative to power law models \cite{Planck13}. The conclusions of the Planck collaboration are based on a uniform model prior as discussed above. With this, the selection of plateau models essentially arises from a phase space volume effect; the fact that there is a larger region of parameter space corresponding to concave models that is not excluded by the data. This leads to a much less pronounced selection of plateau models, without a significant preference for low $r$ models. 
Also, the Planck analysis does not in any way lead to the prediction of a regime of eternal inflation higher up the potential.

\section{Discussion}
\label{disc}

As a quantum mechanical system the universe has a quantum state. A theory of that state acts as a theory of initial conditions in cosmology and is a necessary part of a complete and predictive theoretical framework. We have shown that the quantum state in combination with the structure of the string landscape yields a cosmological measure, or prior, that can to some extent be tested with observations. The prior predicts a probability distribution over cosmological backgrounds, which implies a relative weighting over different landscape regions and hence over observables that differentiate between different regions.

We computed the probabilities for CMB related observables implied by the semiclassical no-boundary wave function in a toy model landscape that contains a range of inflationary patches representative of single-field models with only a few parameters. We found that in this landscape the NBWF predicts our classical cosmological background emerged from a region of eternal inflation associated with a plateau-like potential. The predictions for the fluctuations on observable scales are characteristic of concave potentials, and therefore in excellent agreement with the Planck data. Hence our results disprove the claim in \cite{Steinhardt13} that the Planck observations put pressure on the inflationary paradigm. Having said this, it should also be clear from our work that a truly predictive framework for inflationary cosmology requires a knowledge not only of the structure of the landscape model but also of the specific measure implied by the universe's quantum state. 

This can be illustrated by replacing the NBWF in our formulae with an alternative theory of initial conditions. For instance we can consider theories of initial conditions that strongly favor inflation at high values of the potential, such as the tunneling wave function \cite{Vilenkin86} or chaotic initial conditions \cite{Linde83}. The top-down weighting \eqref{td} is the same for all states. Hence it follows from \eqref{plot} that those theories predict our universe emerges from a region of eternal inflation near the Planck density. In our landscape model this selects power law potentials $\sim \phi^n$, which typically have $n >2$ at high values of the potential. The Planck observations therefore disfavor states like the tunneling wave function in models of this kind\footnote{A specific structure on the landscape that provides a strong bias towards power law patches with low $n$ at intermediate values of $\phi$, e.g. following from a mechanism such as the potential flattening discussed in \cite{Dong11}, may alleviate this problem.}.

\noindent{\bf Acknowledgments:} I thank the participants of the Primordial Cosmology program at the KITP in Santa Barbara for many stimulating discussions, and James Hartle and Mark Srednicki for helpful discussions and comments on a draft of this paper. I also thank the KITP and the Physics Department at UCSB for their hospitality. This research was supported in part by the US National Science Foundation under Grant No. NSF PHY08-55415 and NSF PHY12-05500, by the National Science Foundation of Belgium under the FWO-Odysseus program and by Joe Alibrandi.

\eject

\end{document}